\documentclass[12pt]{article}

\usepackage{arxiv}
\usepackage[utf8]{inputenc} 
\usepackage[T1]{fontenc}    
\usepackage{hyperref}       
\usepackage{url}            

\usepackage{hyperref}          
\usepackage[switch]{lineno}    
\usepackage{amsmath}
\usepackage{booktabs,longtable,array,pdflscape,ragged2e,enumitem}
\newcolumntype{P}[1]{>{\RaggedRight\arraybackslash}p{\dimexpr#1-2\tabcolsep\relax}}
\renewcommand{\arraystretch}{1.2}

\usepackage{booktabs}       
\usepackage{amsfonts}       
\usepackage{nicefrac}       
\usepackage{microtype}      
\usepackage{lipsum}
\usepackage{graphicx}
\graphicspath{ {./images/} }
\usepackage{csquotes}
\usepackage{caption}
\usepackage{makecell}
\usepackage{array}
\usepackage[table]{xcolor}
\usepackage{colortbl}
\usepackage{multirow}
\usepackage{float}
\definecolor{err0}{RGB}{255,255,255}      
\definecolor{err1}{RGB}{230,245,255}      
\definecolor{err2}{RGB}{190,220,250}      
\definecolor{err3}{RGB}{140,190,240}      
\definecolor{err4}{RGB}{70,130,200}       
\definecolor{err5}{RGB}{40,70,120}        

\usepackage{titlesec}
\titlespacing*{\section}{0pt}{*1}{0pt} 

\usepackage{soul}
\usepackage{geometry}
\definecolor{rwblue}{RGB}{210, 235, 255}        
\definecolor{responseblue}{RGB}{190, 215, 255}  
\definecolor{conceptpurple}{RGB}{240, 220, 255} 
\definecolor{selfyellow}{RGB}{255, 255, 200}    
\definecolor{selfgreen}{RGB}{220,255,220} 
\definecolor{processred}{RGB}{255, 200, 200} 
\sethlcolor{highlightyellow} 

\usepackage{natbib}
\usepackage{tabularx}
\usepackage{rotating}
\usepackage{comment}
\usepackage{setspace}
\usepackage{tablefootnote} 

\title{Can we trust LLMs as a tutor for our students? 

Evaluating the Quality of LLM-generated Feedback in Statistics Exams}
\author{
 Markus Herklotz* \\
  Social Data Science and AI Lab\\
  LMU Munich\\
  Munich, Germany \\
  \textit{*Corresponding author} \\
  \texttt{m.herklotz@lmu.de} \\
   \And
  Niklas Ippisch \\
Social Data Science and AI Lab\\
  LMU Munich\\
   Munich, Germany \\
   \texttt{} \\
   \And
  Anna-Carolina Haensch \\
  Social Data Science and AI Lab\\
   LMU Munich\\
   Munich, Germany \\
   \texttt{} \\
 }

\begin{document}

\maketitle

\begin{abstract}\vspace{-0.8\baselineskip}
One of the central challenges for instructors is offering meaningful individual feedback, especially in large courses. Faced with limited time and resources, educators are often forced to rely on generalized feedback, even when more personalized support would be pedagogically valuable. To overcome this limitation, one potential technical solution is to utilize large language models (LLMs).
For an exploratory study using a new platform connected with LLMs, we conducted a LLM-corrected mock exam during the "Introduction to Statistics" lecture at the University of Munich (Germany). The online platform allows instructors to upload exercises along with the correct solutions. Students complete these exercises and receive overall feedback on their results, as well as individualized feedback generated by GPT-4 based on the correct answers provided by the lecturers. 
The resulting dataset comprised task-level information for all participating students, including individual responses and the corresponding LLM-generated feedback. Our systematic analysis revealed that approximately 7 \% of the 2,389 feedback instances contained errors, ranging from minor technical inaccuracies to conceptually misleading explanations. Further, using a combined feedback framework approach, we found that the feedback predominantly focused on explaining why an answer was correct or incorrect, with fewer instances providing deeper conceptual insights, learning strategies or self-regulatory advice. These findings highlight both the potential and the limitations of deploying LLMs as scalable feedback tools in higher education, emphasizing the need for careful quality monitoring and prompt design to maximize their pedagogical value.
\end{abstract}

\textbf{Keywords:} Automated Feedback, Individual Feedback, LLMs, Learning Analytics, Higher Education  

\section{Introduction}
Feedback can be a highly powerful educational tool \citep{wisniewski_power_2020}. However, delivering individualized feedback in large-scale courses with several hundred students presents a considerable challenge for instructors \citep{topali_instructor_2024}. Due to time and resource constraints, instructors might not be able to  deliver personalized feedback on their own at all, resorting to whole-class comments or methods such as peer-assessment \citep{sun_peer_2014} instead. This means many students in high-enrollment classes may not receive the individual feedback by their instructors that they need to improve.
One promising technical approach to address this limitation is the integration of Large Language Models (LLMs) as virtual tutors \citep{liu_llms_2025}, capable of generating personalized feedback at scale. However, despite their potential, LLM tutors might exhibit notable weaknesses, such as the generation of erroneous feedback \citep{qinjin_jia_assessing_2024} and a lack of pedagogical nuance without careful design \citep{dai_can_2023} .

A central consideration for instructors to implement automated technologies in their teaching is the \textit{trustworthiness} of these systems for direct student interaction (\cite{feldman-maggor_impact_2025}, \cite{nazaretsky_instrument_2022}, \cite{viberg_what_2024}, \cite{ayanwale_exploring_2024},  \cite{lyu_understanding_2025}). Trust, in this context, can be understood as an instructor’s willingness to rely on the system and accept its accompanying vulnerabilities, grounded in the belief that generative AI can reliably enhance educational outcomes \citep[p. 2]{lyu_understanding_2025}. Conversely, distrust arises when such systems are perceived as \textit{unreliable, harmful, or misaligned with instructional goals}, prompting educators to withhold reliance or limit use \citep[p. 2]{lyu_understanding_2025}.

\begin{figure}[H]
    \centering
    \includegraphics[width=1\linewidth]{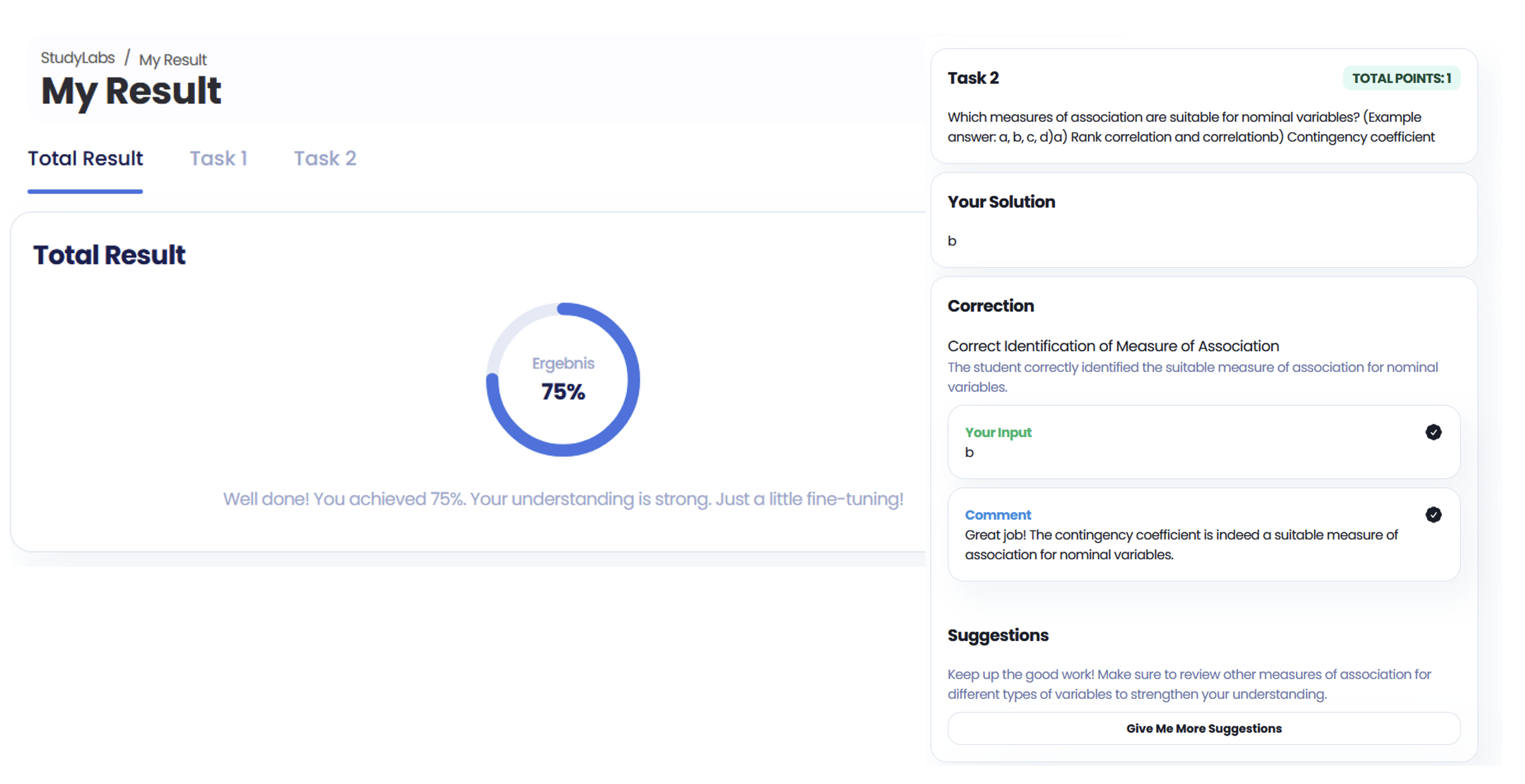}
    \caption{Screenshots of receiving feedback on the \textit{StudyLabs} platform.}
    \label{fig:studylabs}
\end{figure}

In order to leverage the capabilities of LLMs for generating individualized student feedback while minimizing errors, we used the platform \textit{StudyLabs}, on which students received automated, personalized feedback generated by the LLM GPT-4 based on correct solutions provided by instructors (fig. \ref{fig:studylabs}). To generate the feedback, \textit{StudyLabs} submits the students' solutions, scoring guidelines, the correct answers, and the original task descriptions to the OpenAI API. This integration aims to ensure that the feedback provided is consistent with the instructor’s expectations and aligned with the intended learning outcomes of the course. 

Nonetheless, there remains a risk that LLMs may produce misleading, incomplete, or inconsistently structured feedback \citep{li_am_2023}. Prior research has shown that LLM-generated feedback can still contain substantial errors and hallucinations, even when given correct solutions \citep{qinjin_jia_assessing_2024}. Hence, it is necessary to determine whether, and to what extent, the feedback produced under these conditions still contains factual inaccuracies or misleading answers. In addition, reliability extends beyond factual correctness to the structure and consistency of feedback: students who provide identical answers should potentially receive comparable guidance, yet LLMs may respond differently to each case \citep{jacobsen_promises_2023, liang_internal_2024}, potentially creating unequal learning opportunities. Instructors need a clear understanding of the nature, structure, and quality of the feedback that will be provided to their students.

In this study, we systematically analyze LLM-generated feedback from a mock exam in an introductory statistics lecture at a German university. We first identify and classify errors to assess their prevalence and discuss their potential pedagogical impact. We then evaluate the feedback based on established theory, extending the \citet{hattie_power_2007} feedback model by subdividing \textit{task}-level comments into \emph{right/wrong}, \emph{response-oriented}, and \emph{conceptually-focused} \citep{ryan_beyond_2020}, while retaining the original \emph{process}, \emph{self-regulatory}, and \emph{self} levels. We examine how the feedback was usually structured, its variance by task type and performance level and whether students with identical scores receive comparable guidance. Across 2,389 GPT-4 feedback instances from an authentic classroom setting, we quantify error prevalence and propose a combined feedback framework that supports our structured analysis and opens new avenues of designing feedback for future task-based learning environments.

In the \textit{Background} section, we first discuss feedback theories in general before addressing recent studies specifically analyzing LLM-generated feedback. Based on this, we then present our assessment framework and research questions. Following this, we present our methodology, detailing the sample characteristics and the instruments employed for data collection and analysis. In the subsequent \textit{Results} section, we offer a comprehensive analysis of the feedback, incorporating both quantitative and qualitative perspectives. Finally, in the \textit{Discussion and Conclusion} section, we contextualize these findings, draw conclusions, and outline directions for future research.

\section{Background}

\subsection{What is Feedback?}
If used appropriately, feedback can be a highly powerful educational tool \citep{wisniewski_power_2020}. Accordingly, many studies have investigated what constitutes effective feedback and under which circumstances \citep{lipnevich_review_2021}. In their meta-review, \citet{lipnevich_review_2021} analyzed 14 different feedback models and reported that \citet{hattie_power_2007} was both the (by far) most-cited (>,14{,}000 citations at the time) and the only framework unanimously selected by their consulted.

Following this popular model, feedback can be `conceptualized as information provided by an agent (e.g., teacher, peer, book, parent, self, experience) regarding aspects of one's performance or understanding' \citep[p. 81]{hattie_power_2007}. 

Effective feedback addresses three major questions: \textit{Where am I going? How am I going? Where to next?} When there is a discrepancy between what a student understands and what is aimed to be understood, the answers to these feedback questions can enhance learning. In these instances, feedback can help to reduce this discrepancy by raising motivation and facilitating processes that lead to understanding. The model outlined by \citet{hattie_power_2007} differentiates between four levels of feedback that influence its effectiveness: 
\begin{itemize}
    \item  \textit{Task level}: How well a task is performed
    \item   \textit{Process level}: Strategies for the process of solving the task
    \item \textit{Self-regulatory level}: Strategies for self-evaluation and engagement based on already existing knowledge
    \item  \textit{Self level} : Feedback directed to the recipient on a personal level (usually praise)
\end{itemize}

While the self level in \cite{hattie_power_2007} is considered to be rarely effective, the other three levels are interrelated. Effective feedback moves from the task to the process to the self-regulatory level. This ensures that the focus is not on the specifics of the task itself but leads instead to higher engagement with and deeper understanding of the underlying concepts.

Extending this model, recent work of Hattie from the learner’s perspective emphasizes that feedback will only be used when it is heard, understandable, and actionable, and when students judge the “transaction costs” of engaging with it (effort, time, emotional risk) to be worth the anticipated performance gains \citep{mandouit_revisiting_2023}. Purely past-oriented comments or correctness checks therefore have limited educative value unless they include explicit feed-forward and built-in opportunities to apply guidance in subsequent tasks \citep{mandouit_revisiting_2023}. Ideally this is part of an ongoing dialog among students, peers, and teachers that develops feedback literacy, shifting the criterion of effective feedback from what is given to what is actually understood and acted upon by learners \citep{mandouit_revisiting_2023}.

\subsection{Research on LLM feedback}
In recent years, there have been many studies that evaluated LLM-generated feedback, following different approaches and use cases. One main difference in those studies are the various types of exercises they use for their feedback implementations. We can observe two predominant task domains for implementation: elaborative writing tasks such as essays \citep{dai_can_2023, gombert_automated_2024, meyer_using_2024, steiss_comparing_2024, jansen_empirische_2024, gombert_automated_2024, venter_exploring_2025, wan_exploring_2024} and programming exercises \citep{estevez-ayres_evaluation_2024, phung_generative_2023, phung_automating_2024, roest_next-step_2023, hellas_exploring_2023, qinjin_jia_assessing_2024, lohr_youre_2025, er_assessing_2025}. 

While these different types of exercises are quite obvious to classify, the underlying theoretical frameworks and evaluation schemes for assessing LLM-generated feedback remain highly fragmented. Some studies purely focused on the assessment of errors and technical accuracy \citep{estevez-ayres_evaluation_2024, qinjin_jia_assessing_2024}, whereas others prioritize the pedagogical quality and perceived usefulness of the feedback \citep{jansen_empirische_2024, meyer_using_2024, er_assessing_2025, gombert_automated_2024}. Another line of research integrates both dimensions, evaluating error occurrence together with quality criteria often comparing LLM-generated with human feedback  \citep{dai_can_2023, lohr_youre_2025, jurgensmeier_generative_2024, wan_exploring_2024}. Just as the methodological approaches vary widely, so too do the findings reported across the studies. On the one hand, there is evidence that LLM feedback might increase students' performance compared to receiving no feedback \citep{makransky_beyond_2025} and was rated positively by students regarding the tone and scope \citep{roest_next-step_2023, jurgensmeier_generative_2024}. At the same time, students criticize  that LLM feedback lacks depth and helpfulness \citep{jurgensmeier_generative_2024, roest_next-step_2023} and preferred human feedback over LLM feedback \citep{jansen_empirische_2024}. Also in terms of students' performance, there is evidence that human feedback still outperforms LLM feedback \citep{makransky_beyond_2025}. When analyzed by experts, the evidence is contradictory: some find that LLM feedback outperforms human feedback, e.g., in terms of correctness, informativeness, conciseness, and comprehensibility \citep{phung_generative_2023, phung_automating_2024}, or in terms of mentioning the relevant criteria, specificity, and mentioning additional explanations \citep{jacobsen_promises_2023}. Other studies found that LLM feedback is not able to outperform human feedback, for example when assessing if the feedback provided instructions for improvement, was accurate, prioritized essential features, and used a supportive tone \citep{steiss_comparing_2023}. In terms of errors, some studies raise concerns: LLM feedback is often not able to detect the issue \citep{estevez-ayres_evaluation_2024, hellas_exploring_2023} or reports non-existent issues \citep{hellas_exploring_2023}. Experts assessing the feedback in the study by \cite{wan_exploring_2024} found that one third of feedback needs major revisions or needs to be rewritten completely. Overall, the previous research indicates that LLM feedback performs well to some extent, especially when it comes to the style and tone, and might be preferred over no feedback. On the other hand, these new research approaches also suggest moderate to high prevalence of errors as well as concerns whether it can outperform human feedback.

\subsection{A Combined Feedback Framework}
Our central question is whether teachers can rely on LLMs with providing feedback to their students. Hence, we must weigh accuracy and instructional value in equal measure. The relevant counterfactual in a large-enrollment course is no individual feedback at all, so we will not benchmark against human feedback. Instead, we judge the LLM output against established feedback theory, most prominently the four-level model of \citet{hattie_power_2007}, still the field’s dominant framework \citep{lipnevich_review_2021}.

Empirical work shows a persistent skew toward the task level in both human and AI feedback (for example recently by \citet{dai_can_2023}). Process, self-regulatory and self feedback typically account for a very small amount of feedback. Given this consistent trend, we anticipate our LLM to produce predominantly task-level comments as well. To yield deeper analytic and qualitative insight, we further dissect the task-level category. For this purpose, we draw on the feedback taxonomy by \citep{ryan_beyond_2020}, which we expect to comprehensively capture the spectrum of task-level feedback, especially for our statistics mock exam:
\begin{enumerate}
    \item \textit{Right/wrong feedback}: Simple identification of the correct response
    \item \textit{Response-oriented feedback}: Short explanation of why each option was correct or incorrect
    \item \textit{Conceptually-focused feedback}: Discussion of the correct response in more detail (e.g., the underlying concepts)
\end{enumerate}

In their study, \citep{ryan_beyond_2020} employed different tasks related to each other to analyze the effect of these kinds of feedback on the near- and far-transfer of knowledge. They found "response-oriented" feedback and "conceptually-focused" feedback superior to simple "right/wrong" feedback.

For the evaluation of the LLM generated feedback, we will combine the approaches by \citet{hattie_power_2007} and \citet{ryan_beyond_2020}. We will dissect the task level in the three categories: "right/wrong," "response-oriented," and "conceptually-focused" based on \citet{ryan_beyond_2020}. This refinement also aligns with more recent developments in the Hattie framework, which from a learner's perspective emphasizes that purely corrective comments have limited educative value \citep[see Section~2.1]{mandouit_revisiting_2023}. Accordingly, contrasting the three task-level categories from simple "right/wrong" comments to future-oriented "conceptually-oriented" guidance should yield clearer distinctions and finer-grained insights regarding into the structure and pedagogical value of the feedback. The other \cite{hattie_power_2007} levels remain as they are. Accordingly, we define the feedback categories and their guiding questions for our analysis based on the classfications as follows:

\begin{enumerate}
    \item \textbf{\textit{Task level}}: 
    \begin{enumerate}
        \item \textit{\textbf{Right/wrong}}: Does the feedback text indicate whether the answer is correct or not?
        \item \textit{\textbf{Response-oriented}}: Does the feedback text explain why the answer is correct or not?
        \item \textit{\textbf{Conceptually-focused}}: Does the feedback text provide an explanation that enhances understanding of the underlying concept of the task?
    \end{enumerate}
\item \textbf{Process level}: Does the feedback text provide strategies to facilitate the student's learning?
\item \textbf{Self-Regulatory level}: Does the feedback text provide suggestions on how the student can control and manage their own learning?
\item  \textbf{Self level}: Does the feedback text evaluate the student on a personal level?
\end{enumerate}

\subsection{Research Questions}

 The present study investigates the quality and structure of feedback generated by LLMs in educational contexts. Specifically, it examines the occurrence of errors in feedback when correct solutions are available, as well as the structural characteristics of the feedback produced. The analysis is guided by two overarching research questions: (1) To what extent does the LLM produce errors while giving feedback when the instructor provides the correct answers? and (2) What is the structure of the feedback generated, including its variations across tasks and students, and its alignment with established educational feedback theories?

\section{Data \& Methodology}

In our analysis, we rely on three instruments: Data gathered from an online survey both at the start (1) and at the end (2) of the course, as well as the passively tracked data from the mock exam on \textit{StudyLabs} (3). The sample and the methods used for the analysis will be outlined below.

\subsection{Sample of students}

Our study subjects are Bachelor students who participated in the 'Introduction to Statistics' lecture at the LMU Munich during the winter term 2024. This is a large course of about 300 students\footnote{This is about the number of students who take the exam every year. In-class participation is not mandatory. Therefore, it is not possible to provide exact numbers of participants.}, of whom 202 completed the entry survey in the first week of the term. The course is mandatory either for students with a major in sociology or (media) informatics, or for students with 'Statistics and Data science' as their minor. The lecture was held in person and streamed online; both the weekly exercise and tutorial sessions were conducted in person only. The course was held entirely in German. Both surveys were programmed online but fielded in person during the lecture to raise response rates. The entry survey reveals that most students in this course are enrolled as Bachelor Sociology majors (75.12\%). Among the other majors, Media Informatics (10.4\%) and Geography (3.5\%) are the most prominent. Regarding the minor study subjects of the students, the biggest group is students with Statistics as their minor (19.7\%). The students in this course primarily identify as female (68.3\%) and very early in their studies, mainly in their first term (85.6\%); some are in the third semester (12.4\%). 

\subsection{Mock exam and student survey}

For gathering insight on the feedback given by the LLM on the \textit{StudyLabs} platform, one of the last sessions of the course in January 2024 was used to host a mock exam in person. The mock exam was used one year prior as the actual exam and was conducted in German. The students took the mock exam on \textit{StudyLabs}\footnote{We deployed \textit{StudyLabs} using version GPT-4-0613 without any changes to the \textit{StudyLabs} system prompt to achieve an authentic classroom setting: "\textit{As an AI, you are designed to support students with their exam questions, based on a collection of corrected tasks. Your role is to provide clear, supportive, and constructive answers by integrating the context of the task, model solutions, student responses, assessment criteria, and previous feedback. Use the chat history to maintain continuity in the dialogue and ensure that your responses dynamically align with the evolving conversation. Present mathematical content in KaTeX format and consistently use positive, motivating language. Tailor your explanations to the student's level of understanding and use relevant examples or analogies to clarify complex concepts. Your interaction should be professional, pedagogically sound, and strictly focused on the provided material and student inputs.}"} entirely and received the LLM-generated feedback after requesting it (usually after the last exam question). From this mock exam on \textit{StudyLabs}, \textit{ZAVI} shared the data with the team of authors. The data set contained all questions asked in the mock exam, the achieved and achievable points per exercise, the student answers, and the LLM-generated feedback for all 38 exercises anonymized with a randomly generated username.

In total, we have data from 70 students and their mock exams. Most of the students worked through all exercises, yielding a total of 2,389 exercises and feedback instances. To structure the exam for the analysis, we categorized the 38 exercises into four different types of tasks (knowledge, interpretation, calculation, and R) and eight different categories of statistical concepts (measures of central tendency, measures of dispersion, measures of correlation, visualization, regression, variables, frequencies, and R).\footnote{The english translation of the exam including the categorization in task types is available as supplementary material.} 

The in-depth evaluation and coding of the feedback was done by two of the authors. First, one author that is also responsible for the grading of the actual exam of the course, screened all of the 2,389 feedback instances individually for potential errors, yielding a binary variable indicating an error or not. At the same time, we stored the feedback instances with the visually marked errors separately for content analysis. Here, we chose a rather broad definition of error and decided to flag the feedback that might be misleading or confusing for students in their first semester and first statistics course (see also section 4.2). The flagged errors were subsequently checked by another member of the author team. Eight differences between the authors arose, were discussed and resolved, yielding five changes in the flagging.

To assess the feedback structure in detail based on our feedback categories (section 2.3), we chose a subset of the 38 mock exam tasks to be analyzed with a qualitative content analysis. We selected one question from each task category (knowledge/calculation/interpretation) to cover the didactic range of exercises. To allow comparisons within the same task category, we additionally chose a second interpretation question for a total of four exam tasks to be analyzed in detail. 

The first question is a multiple choice \textit{knowledge} question, where students were expected to choose measures of central tendency (a) Mean, b) Median, c) Variance, d) Mode). All participating students except one provided an answer and received corresponding feedback (n = 69). Secondly, we chose an \textit{interpretation} question where students were supposed to argue, based on a histogram and boxplot, whether they expected the mean or the median to be higher. All participating students received feedback here (n = 70). Lastly, feedback on a \textit{calculation} exercise was analyzed (n = 69). Here, variances of weight based on five measurements separately for two groups needed to be calculated.

 As a method, a qualitative content analysis \citep{mayring_qualitative_2014} was chosen, in which the feedback categories outlined above were used as categories for coding. The coding was conducted by two of the authors. Before the analysis, a part of one of the tasks was coded independently and then discussed, both to test the coding scheme and to establish inter-coding reliability. Afterwards, the authors split the tasks in half (even and uneven student ID numbers). After the procedure, both authors checked the other persons' coding for possible disagreements, but no were identified. Every feedback text part was coded only with one category. All text passages were coded. Hence, adding up all coded passages yield exactly 100\% of feedback , allowing a comparison of the coverage of different categories. Since the provided feedback was in German, we translated examples for the purpose of this paper.

To capture the students' experience, an online survey was designed for the start and the end of the course. The survey was conducted using in-class time to raise participation. Afterwards, the link was also shared on the online learning platform moodle available for everyone that could not attend class. The survey mainly consisted of questions addressing the students' socio-demographic information, previous usage of LLMs and experience with the \textit{StudyLabs} platform.

\section{Results}

\subsection{Students' experience}

In the context of introducing an AI tool for students, their experience and perspective is an important part for evaluating the tool. Additional to the evaluation of the \textit{StudyLabs} platform, we asked for the existing experience with ChatGPT\footnote{(Chat)GPT is not the only LLM which applies here, but for example due to media coverage, the most likely for students to have had contact with. We decided that a question detailing other LLMs could have added more confusion among students than the potential benefit of insight.}. The majority of respondents of the start-of-semester survey (N=202) stated that they have used ChatGPT already (54\%), and 5\% stated that they do not know what ChatGPT is. Of those who said to have used it before, the frequency of usage is very heterogeneous.
Regarding the range of use, 43\% stated that they use it for explanations, 34\% for the production of text, and 23\% for programming (multiple answers possible). In the end-of-semester survey (N = 104), 65.8\% stated that they participated in the mock exam with \textit{StudyLabs}. Among those, most have used ChatGPT already before the mock exam (67 \%) with approximately three-quarters of them very frequently. 

Of those who stated to have participated at the mock exam and to have used ChatGPT before (n = 32), the usage of ChatGPT for explanations (97 \%) was most prominent, whereas usage for production of text and programming (both 34 \%) was used less (multiple answers possible). 

Overall, the feedback comments of the LLM provided on the \textit{StudyLabs}-Platform were rated useful (93 \% rather or very useful), and respondents would recommend it (78 \% would definitely or rather recommend). One reason for not recommending might be the interface for submitting solutions, which was rated rather low (37 \% stated that it was (rather) not user-friendly.\footnote{One potential reason for this rating could be that students were asked to use LaTeX code for submitting formulas, but they were not very familiar with LaTeX yet.}. Interestingly, no difference in the perceived usefulness was found based on whether students already had used ChatGPT before or how frequently they used it.

\subsection{Errors in the LLM-generated feedback}

Out of the 2,389 feedback instances screened by two coders that are also authors of the paper, 6.99\% (\textit{n} = 167) were identified to contain at least one error (see table \ref{tab:correctness}). The errors occurred mainly in feedback for incorrect student answers (\textit{n} = 137) but to a lesser extent also for correct student answers (\textit{n} = 30).

\begin{table}[h]
    \centering
    \renewcommand{\arraystretch}{1.5}
    \begin{tabular}{l | c c | c}
        & \multicolumn{2}{c|}{\textbf{Student Answers}} & \\
        \textbf{Feedback} & Correct & Incorrect & Total \\
        \Xhline{1pt}
        No errors & 46.92\% & 46.09\% & 93.01\% \\
        Errors    & 1.26\%  & 5.73\%  & 6.99\%  \\
        \Xhline{1pt}
        Total     & 48.18\% & 51.82\% & 2,389  \\
    \end{tabular}
    \caption{Feedback instances containing (no) errors by correct and incorrect student answers.}
    \label{tab:correctness}
\end{table}

We observed erroneous feedback instances across various topics and task types in the mock exam, without a consistent pattern (see Table \ref{tab:error-tasks}). For 13 out of the 38 tasks, the LLM produced only correct responses. The majority of tasks (19) contained between 1 and 5 erroneous feedback instances.\footnote{Erroneous feedback instances are responses that contain at least one error, without counting multiple errors within a single response more than once.} Four tasks resulted in 6 to 10 erroneous feedback instances. Notably, both tasks with 11 to 15 erroneous feedback instances were correlation-related, but drawn from different sections of the exam and pertaining different task types. Two tasks, covering different topics and task types, stood out with much higher error frequencies of over 30 erroneous feedback instances.

 These erroneous feedback instances range from individual words being wrong to technical difficulties accepting correct answers to plain wrong explanations of statistical concepts. Instead of categorizing these errors in a broader typology and describing them abstractly, we will first examine them in more qualitative detail in the following section. This detail seems necessary to generate insight into \textit{what} can go wrong and to conceptualize approaches for further reducing errors in the future (see Table \ref{tab:errors-most-common} in appendix with most common errors and examples). 

\renewcommand{\arraystretch}{1.4} 
\setlength{\tabcolsep}{10pt}     

\begin{table}[ht]
\centering
\caption{Error distribution over topics and task types.}
\begin{tabular}{l!{\vrule width 1.2pt}c|c|c|c}
\textbf{Topics} & \multicolumn{4}{c}{\textbf{Task types}} \\
\textbf{} & \textbf{Knowledge} & \textbf{Calculation} & \textbf{Interpretation} & \textbf{R} \\
\Xhline{1.2pt}
\textbf{Central Tendency} 
  & \cellcolor{err0}1a & \cellcolor{err0}3b & \cellcolor{err1}2b &  \\
  & \cellcolor{err0}1h & \cellcolor{err2}3c &                    &  \\
  & \cellcolor{err5}3f & \cellcolor{err1}3d &                    &  \\
  &                    & \cellcolor{err1}3e &                    &  \\
\hline
\textbf{Dispersion} 
  & \cellcolor{err1}1b & \cellcolor{err1}4a &                    &  \\
  &                    & \cellcolor{err1}4b &                    &  \\
\hline
\textbf{Correlation} 
  & \cellcolor{err1}1c & \cellcolor{err0}5a & \cellcolor{err0}5b &  \\
  & \cellcolor{err3}1e & \cellcolor{err3}6b & \cellcolor{err1}5c &  \\
  &                    & \cellcolor{err1}6c & \cellcolor{err2}6c &  \\
  &                    & \cellcolor{err0}6d & \cellcolor{err0}7d &  \\
\hline
\textbf{Visualization} 
  & \cellcolor{err1}1d &                    & \cellcolor{err1}2a &  \\
  &                    &                    & \cellcolor{err1}2c &  \\
  &                    &                    & \cellcolor{err1}2d &  \\
  &                    &                    & \cellcolor{err0}2e &  \\
\hline
\textbf{Regression} 
  & \cellcolor{err1}1f & \cellcolor{err1}7b & \cellcolor{err2}7a &  \\
  & \cellcolor{err1}1g & \cellcolor{err2}7c & \cellcolor{err0}7d &  \\
  &                    &                    & \cellcolor{err0}7e &  \\
  &                    &                    & \cellcolor{err0}7f &  \\
  &                    &                    & \cellcolor{err1}7g &  \\
  &                    &                    & \cellcolor{err4}7h &  \\
\hline
\textbf{Variables} 
  &                    & \cellcolor{err1}3a &                    &  \\
\hline
\textbf{Frequencies} 
  &                    & \cellcolor{err0}6a &                    &  \\
\hline
\textbf{R} 
  &                    &                    &                    & \cellcolor{err0}2f \\
  &                    &                    &                    & \cellcolor{err0}4c \\
\end{tabular}

\vspace{0.8em}
\textbf{Number of erroneous feedback instances}
\vspace{0.3em}
\begin{tabular}{c c c c c c}
\cellcolor{err0} 0 &
\cellcolor{err1} 1--5 &
\cellcolor{err2} 6--10 &
\cellcolor{err3} 11--15 &
\cellcolor{err4} 31 &
\cellcolor{err5} 34 \\
\end{tabular}
\label{tab:error-tasks}
\end{table}

One example of a technical mistake pertains to question (7a), where students have to interpret a regression coefficient of 0.3021 for the relation between household income and vacation spending. In 10 feedback instances, students answered correctly that this means that per additional Euro of household income, people spend on average \textit{0.3021 Euro }more on vacations. But in these cases, the LLM-generated feedback stated that this is incorrect and that the actual number is \textit{30 cent}. This is very likely due to the provided template solution also noting "30 cents", but neither in the task itself nor in the solution was the criteria for transferring the unit defined. 

The most common error was a systematic mistake, which occurred in almost half of the feedback instances (\textit{n} = 34). For this task (3f), where students had to enter the values of the 25 and 75\% quartile, a new table was introduced in the previous exercise (3e), but the LLM still referred to the information given at the start of this set of exercises (3a - 3d). Hence, it returned plain wrong information when assigning the quartiles. Another example of not using the correct input was in subsequent tasks where the question description provided an intermediate value in case the student did not solve the previous question. In several instances, the LLM ignored the possibility of this provided intermediate value and just assumed the student's answer was incorrect.

In another common feedback error (\textit{n} = 30), the LLM assumed the wrong scale for the independent variable in a regression. In this task (7h), the variable 'children in the household' was defined as binary (1 = yes, one or more children, 0 = no children). When returning feedback, the LLM treated this variable as a metric instead of the number of children in a household, resulting in wrong information. 

Another confusion occurred repeatedly (\textit{n} = 14) in a question (6b) where students had to calculate Chi². The culprits here are the two different, valid approaches to calculating Chi²: Either the 'traditional' way with first calculating the expected values \textit{or} the faster formula, in which you do not have first to calculate expected values. Both approaches were introduced in the lecture in this course, but the formula without expected values was recommended. Hence, the template solution for this mock exam task also included this faster formula. When the LLM returned feedback for this task, it also showed exactly this formula from the template solution but simultaneously explained the method by first calculating expected values instead. Subsequently, students received an explanation of a formula different from the one shown. 

Some other common errors were wrong or insufficient explanations of statistical concepts. This comprises six feedback instances when the LLM only differed between categorical and metric variable scales, using the Spearman and Pearson coefficients evenly while sparing out the information that Spearman is a valid measure for ordinal variables. Further, eight feedback instances showed misleading explanations of an ordinal variable, confusing it with a purely nominal variable.

Further, we also found some minor issues, which are quite heterogeneous and deserve a mention to show the range of feedback errors. In very few (less than five) instances, there were some hallucinations. For example, in a question with a contingency table on the ownership of dogs or cats, the LLM invented birds and fishes in its feedback. In another question, it completely invented a sentence, which we could not decipher where it came from or what its meaning is supposed to be.\footnote{"With that, you have correctly responded to the survey and accordingly coordinated with the disability support services."} As this mock exam and its feedback was conducted in German, in several instances, the translation seemed to go wrong in parts on individual words\footnote{"Datenset" instead of "Datensatz", "Scala" instead of "Skala", "Modeswert" instead of "Modalwert", "Nomininaldaten" instead of "Nominaldaten", "dealen" instead of "zu tun haben".} or it just used the English word entirely\footnote{"descriptive" instead of "deskriptive", "indeed" instead of "in der Tat"} instead of the German word.

As introduced before, these errors are quite heterogeneous in their types. But they might be distinguishable in two categories based on their potential effect on the students: 1. (More) easily identifiable errors, which mainly cause frustration or distraction, and 2. hardly identifiable errors with convincingly formulated but wrong explanations, which 'silently' introduce incorrect concepts into the students' minds. 

In category 1 are the rather technical mistakes, such as not accepting the currency (0.30 Euro vs. 30 cents), ignoring the input of the tasks and the translation, or single word mistakes. Most students should notice quickly that they provided the correct answer when they entered 0.30€ instead of 30 cents. But first of all, it is a frustrating experience to provide the correct answer and not receive the correct grade. Second of all, it could still be misleading for students who were insecure about solving the exercise. While one could argue that the instances of single-word mistakes and translations are minor and should be easily discernible for students, they still can be a distraction in something that is supposed to be a help for them. Apart from disruptive moments just by the nature of encountering errors, the students could wonder if something else is meant and spend valuable time dissecting this hallucinated issue.

While Category 1 can cause specific issues, we would assume that the errors of Category 2 have a more concerning (long-term) impact. As long as the students confidently notice the errors, spending more time on the question and self-enforcing their knowledge could even strengthen their learning. On the contrary, when the feedback explained statistical terms and concepts, at least misleadingly or even plainly wrongly, this could negatively impact the students' understanding. Due to the formulative capacities of LLMs the explanations still \textit{sound} correctly and students are probably not used to consider receiving erroneous information in the classroom. Often, these erroneous explanations were also not plain wrong but insufficient based on the context of the question and the overall learning goals. In these instances, the course instructor would just explain differently or add additional information.

\subsection{Structure of the LLM generated feedback}
Following the identification of errors in the feedback, we investigate in more detail how the feedback is composed. For doing so, we selected four questions that represent the range of tasks in this (mock) exam and coded the corresponding feedback based on our feedback categories derived from \cite{hattie_power_2007} and \cite{ryan_beyond_2020} (see subsection 2.3). As formulated in our research questions, we are especially interested in how the feedback does typically look like, how it does differ between tasks and students, and to which extent it aligns with educational feedback theories.

\subsubsection{Overall Feedback Structure}
\begin{figure}
    \centering    \includegraphics[width=1\linewidth]{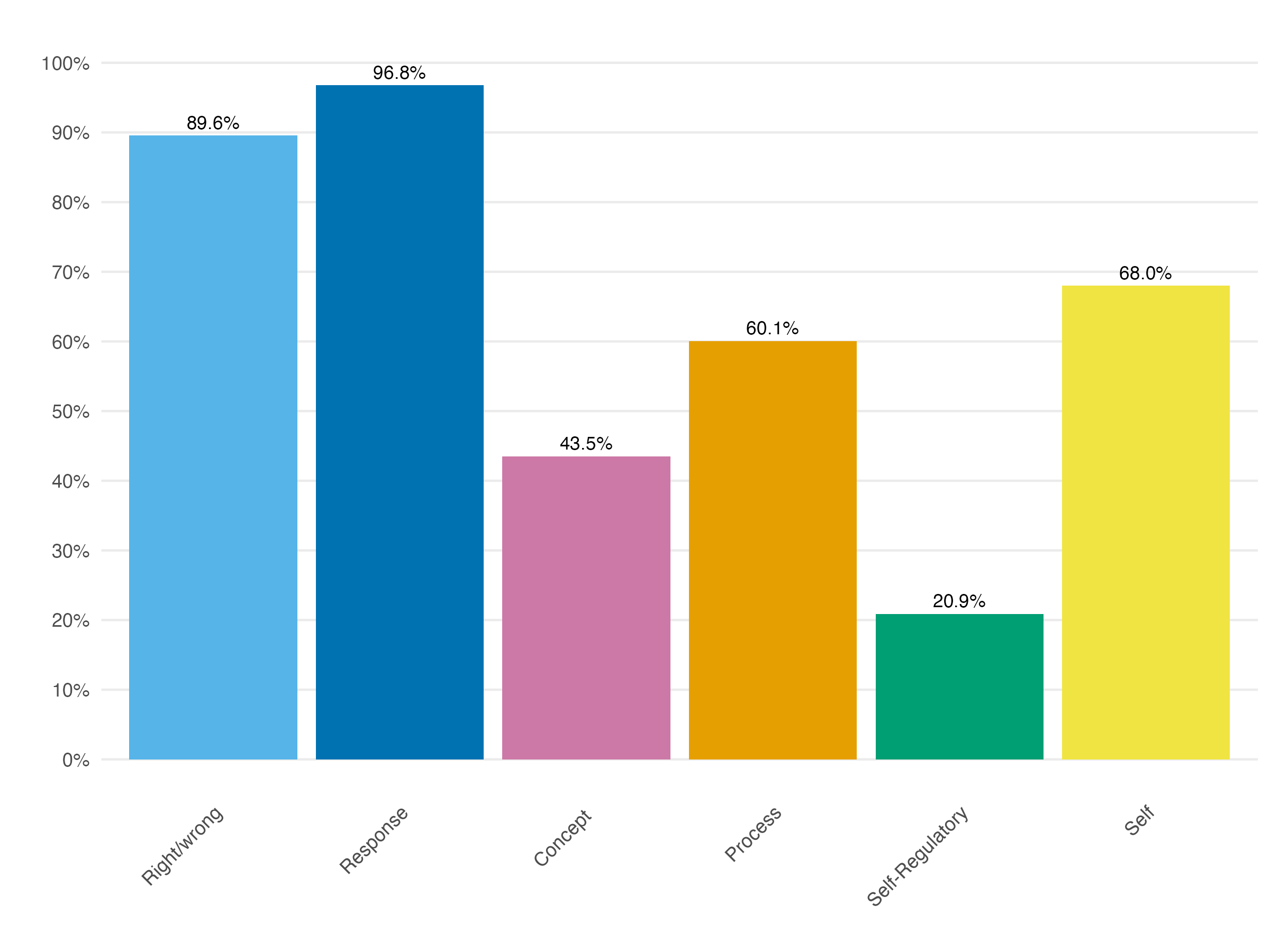}
    \caption{Frequency of feedback categories across all 70 students and 4 tasks.}
    \label{fig:freq-all}
\end{figure}

Overall, across all 70 students and all four tasks, the most common feedback type is "response-oriented" (see fig. \ref{fig:freq-all}) which focuses on explaining why an answer is correct or incorrect, rather than merely stating its correctness ("right/wrong") or providing broader conceptual explanations ("conceptually-focused"). "Response-oriented" feedback occurs in 97 \% of the 276 feedback instances that we coded manually -- hence, in almost all feedback occurrences, bar a few exceptions. “Right/wrong” feedback, which simply indicates whether an answer is correct or not, is also very common (90 \%), appearing in nearly nine out of ten feedback instances. "Self" feedback, which refers to personal evaluations or affirmations directed at the student (such as "Well done!", "Good job!" or "Great!") is present in more than two thirds of the feedback instances (68 \%). "Process" (60 \%) feedback, , i.e. , offering strategies and approaches for learning, still occurs in more than half of the instances. “Conceptually-focused” feedback, which provides additional explanations to deepen understanding, appears less frequently at 44 \%. The - by far - least common feedback type is "self-regulatory" (21 \%), which supports students in independently monitoring and correcting their own learning processes.

\begin{figure}
    \centering
    \includegraphics[width=1\linewidth]{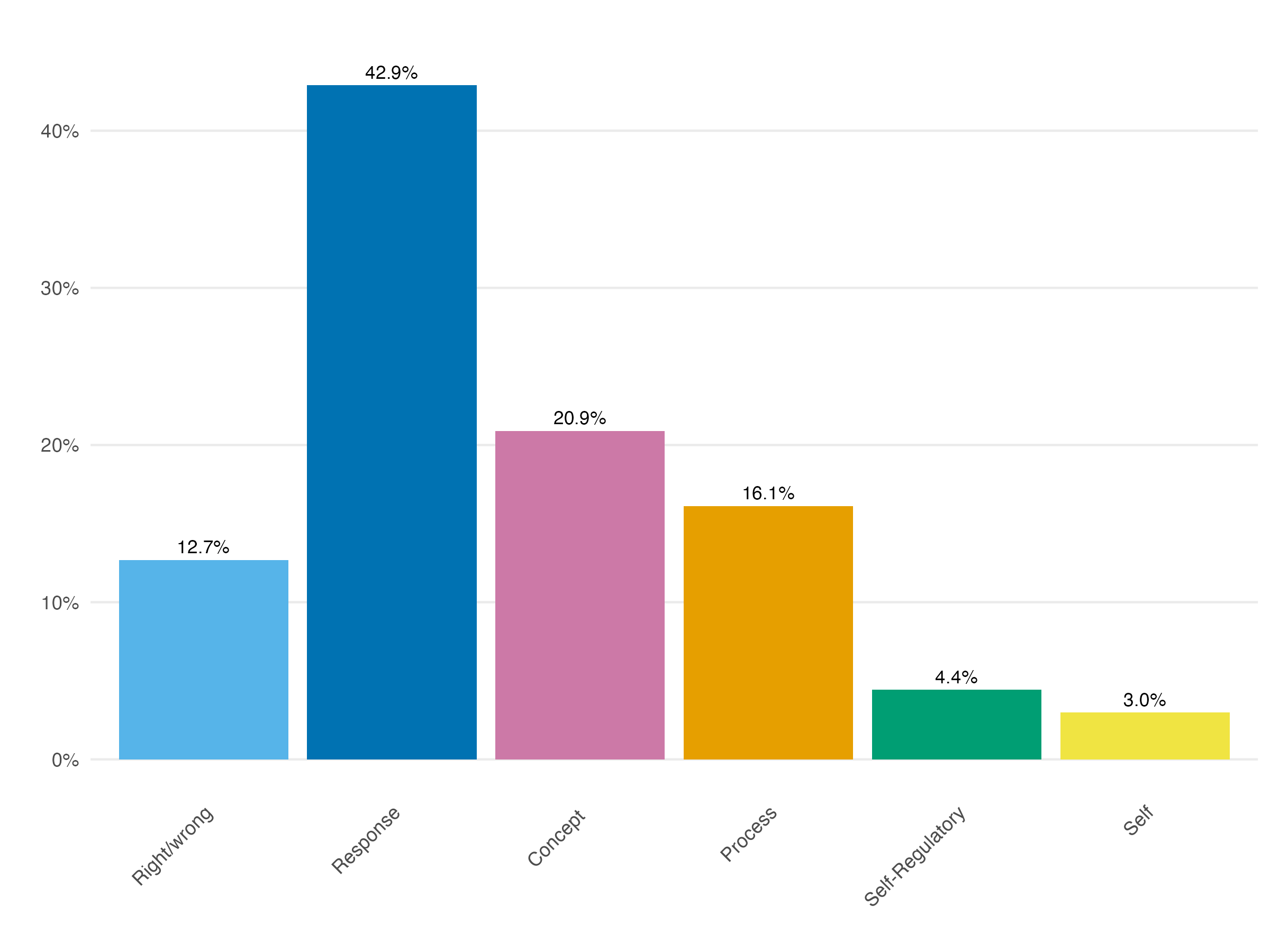}
    \caption{Coverage of feedback categories across all 70 students and 4 tasks.}
    \label{fig:coverage-all}
\end{figure}

In addition to the frequency of feedback categories across all instances, we are also interested in the proportional volume of each category. Some types of feedback may occur in many instances, but only take up a small part of the output. Potentially, shorter feedback types might be perceived by students as less important and thus exert less impact on their learning. To investigate this aspect, we analyze  the number of characters (including spaces) associated with each coded feedback segment. "Response-oriented" is not just the most frequent feedback across the instances, it also accounts for the largest content share of all coded feedback. 43 \% of the provided feedback belongs to this category (see fig. \ref{fig:coverage-all}). "Conceptually-focused" feedback only comprises 21 \% of the feedback coverage, "process" feedback 16 \% and "right/wrong" feedback 13 \%. "Self-regulatory" (4 \%) and "self" (3 \%) are the categories with the lowest feedback coverage. Interestingly, this suggests that merely considering the frequency of feedback categories is insufficient to fully understand how the feedback is composed and how it might be perceived by students. As expected, "right-wrong" is far more frequent than it is high in volume, as it takes not many words to state if an answer is correct or not.
"Process" feedback only occurs in 60 \% of the feedback instances, but takes the third-most overall space of all the feedback categories, indicating longer sequences of this type.

\subsubsection{Feedback Structure by Task}
We selected four tasks that capture the range of the mock exam and examined whether there are differences in the feedback structure between tasks: 
\begin{itemize}
    \item 1a: A multiple choice knowledge question about measures of central tendency.
    \item 2d: A graph interpretation question where students compare median and arithmetic mean.
    \item 3e: A table interpretation question where students have to decide between  median and arithmetic mean based on outliers.
    \item 4a: A table calculation question where students have to calculate two sets of variances and standard deviation.  
\end{itemize}
\begin{figure}
    \centering
    \includegraphics[width=1\linewidth]{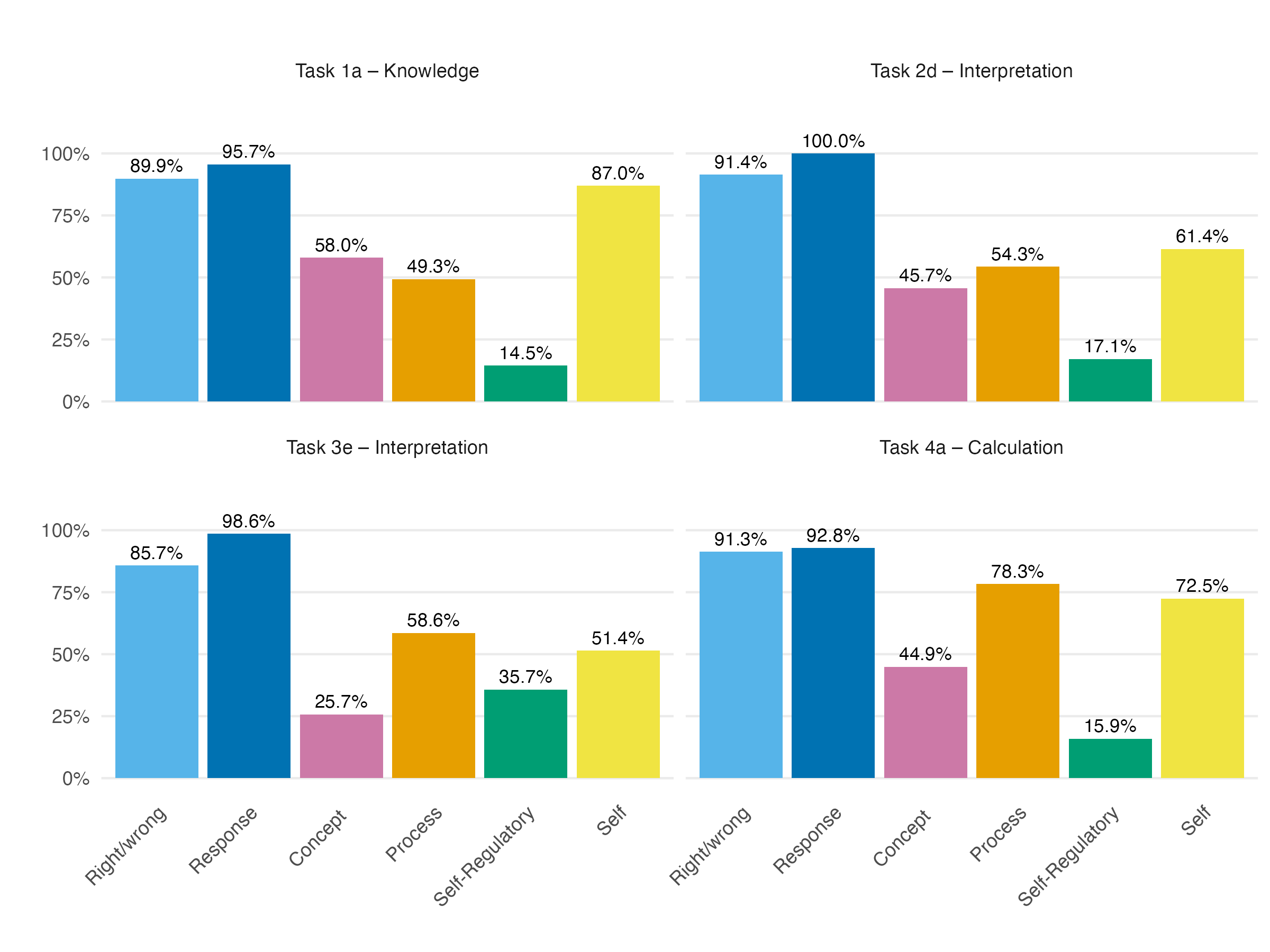}
    \caption{Frequency of feedback categories by task type.}
    \label{fig:freq-task}
\end{figure}

"Response-oriented" feedback is consistently the most frequent category across all 4 tasks and "right-wrong" feedback the second-highest (see fig. \ref{fig:freq-task}). Only in task 4 does ""esponse-oriented" feedback fall under 95 \%, which is also when "right-wrong" is the closest in frequency with only a difference of 1.5 percentage points. 

Regarding the other feedback categories, we can observe stronger differences between tasks. Notably, "self-regulatory" feedback occurs in a very similar frequency of 15 - 17 \% for three of the four tasks, but is much more common in one of the interpretation tasks (3e, 36 \%). At the same time, "Conceptually-focused" feedback is considerably less frequent in this task (25 \%) than in others (46 \% - 58\%). "Process" feedback, on the other hand, is notably more frequent for the calculation task (4a, 78 \% vs. 49 - 59 \%). Feedback of the "self" category is less prevalent for the interpretation tasks (2d: 61 \%, 3e: 51 \%) than for the knowledge (1a: 87 \%) and calculation task (73 \%).

\begin{figure}
    \centering
    \includegraphics[width=1\linewidth]{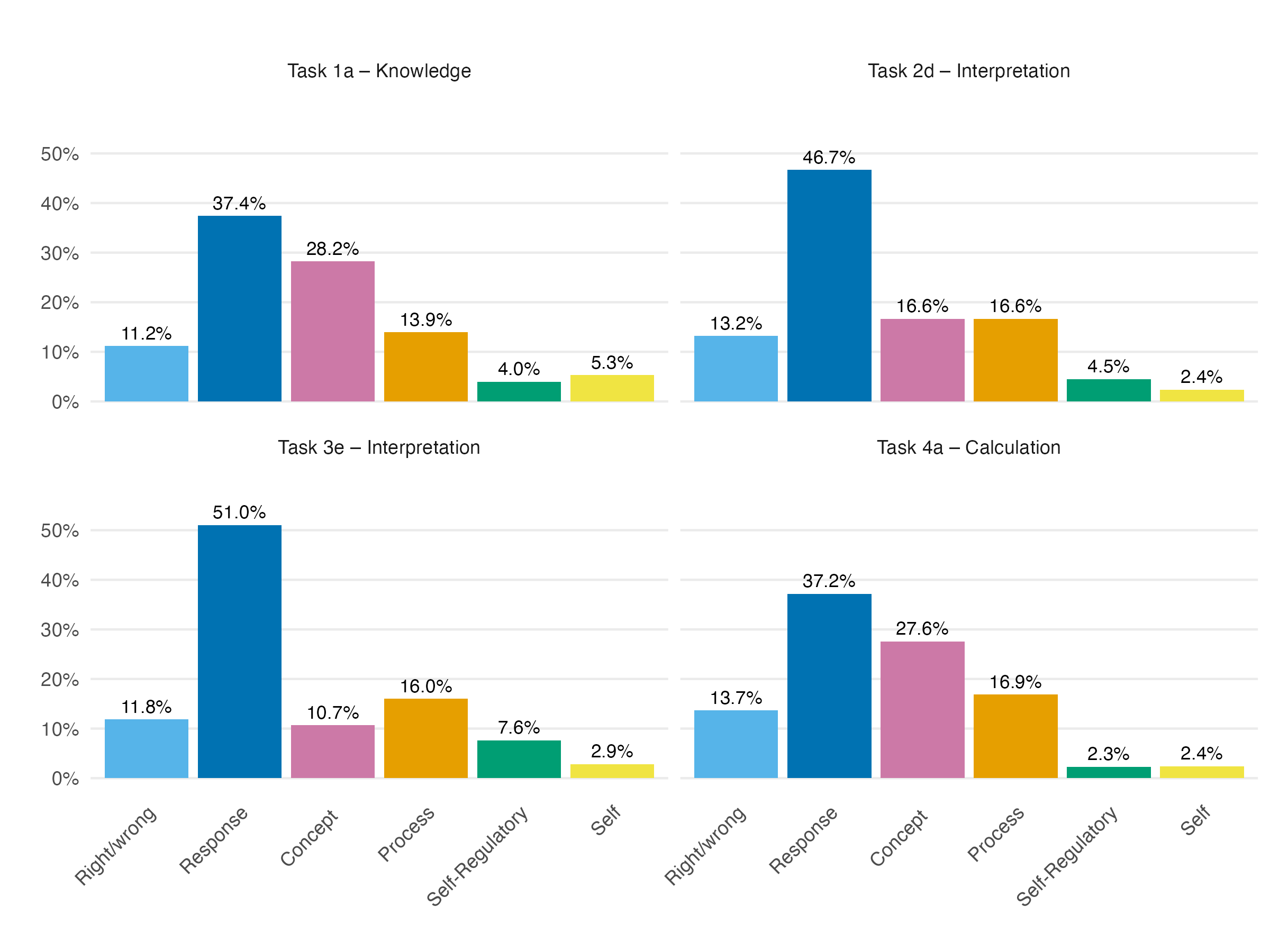}
    \caption{Coverage of feedback categories by task type.}
    \label{fig:cover-task}
\end{figure}

Again, we supplement these findings on frequency by assessing the coverage of each category of feedback across the four tasks (see fig. \ref{fig:cover-task}). As above, the tasks show quite a similar coverage regarding the "right/wrong" category. However, there are bigger differences with response-focused feedback. The interpretation tasks 2d and 3e show a notably bigger coverage of "response-oriented" feedback. Whereas the coverage of process feedback is approximately equal across the tasks, we can see stronger variation with "conceptually-focused" and "self-regulatory" feedback: "conceptually-focused" feedback has a higher coverage for the knowledge and calculation task, while "self-regulatory" feedback is lower for those two compared to the interpretation tasks.

In other words: the knowledge task has a comparatively lower coverage of "response-oriented" and "process" feedback, but the highest for "conceptually-focused" and "self". The interpretation tasks have the highest coverage for "response-oriented" feedback, while the calculation task has the lowest coverage for "self-regulatory" feedback and low coverage for "response-oriented".

\subsubsection{Feedback structure by different points achieved}
We are also interested in examining how the feedback differed between students. A key factor to differentiate students is the number of points they achieved on the respective task. The occurrence and coverage of different feedback types may vary depending on how well the student performed -- similar to how a human instructor would provide different feedback to a student who answered incorrectly compared to one who gave a fully correct answer.

\begin{figure}
    \centering
    \includegraphics[width=1\linewidth]{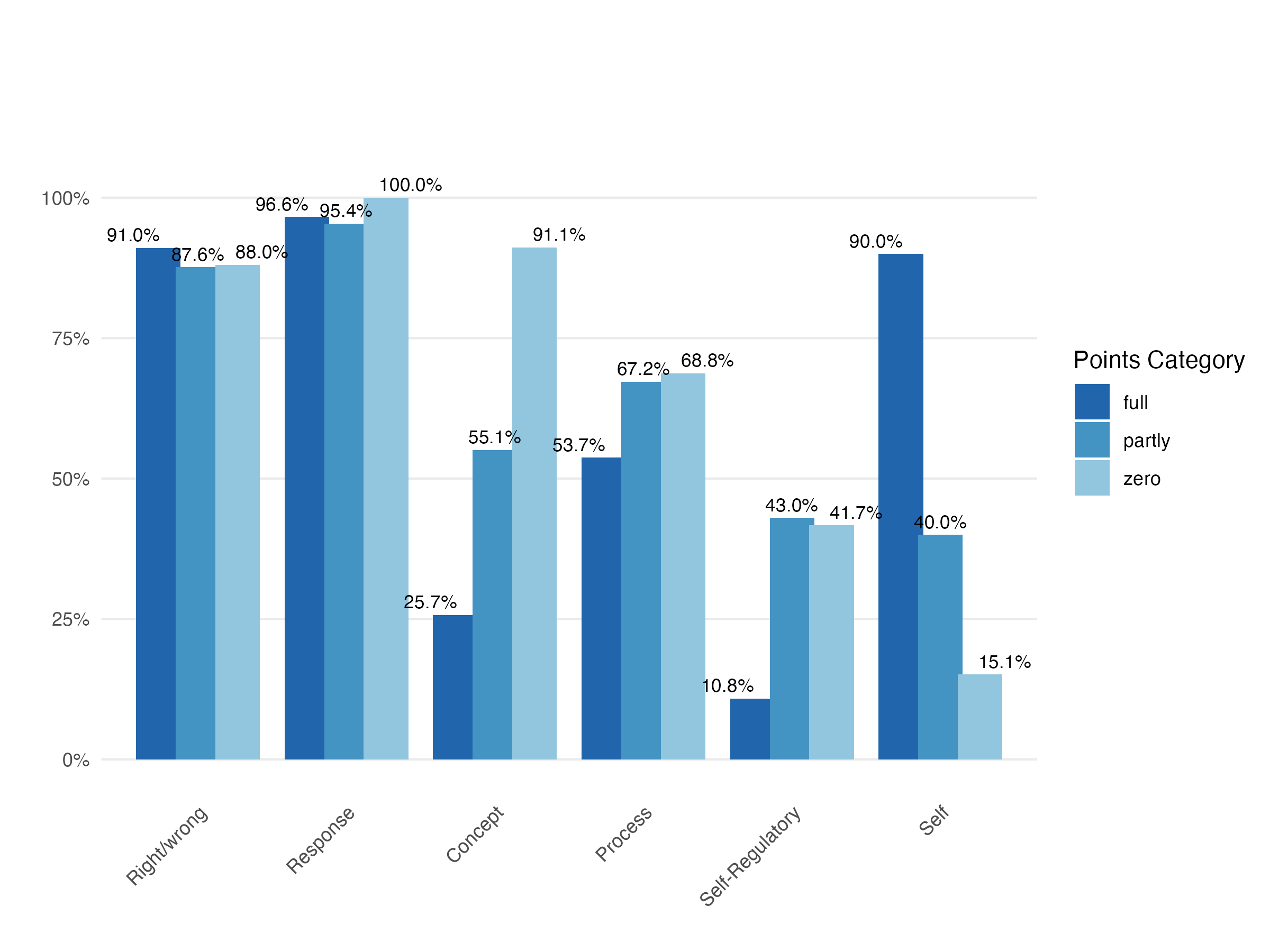}
    \caption{Frequency of feedback categories by points achieved.}
    \label{fig:freq-points}
\end{figure}

Figure \ref{fig:freq-points} indicates that for "right/wrong" and "response-oriented" feedback, the occurrence does not vary meaningfulyl. However, the difference is very big for "conceptually-focused", "self-regulatory", and "self" feedback. Students, who achieved full points received considerably less often "conceptually-focused" and "self-regulatory" feedback and far more often "self" feedback; the opposite is true for students, who achieved zero points. 

\begin{figure}
    \centering
    \includegraphics[width=1\linewidth]{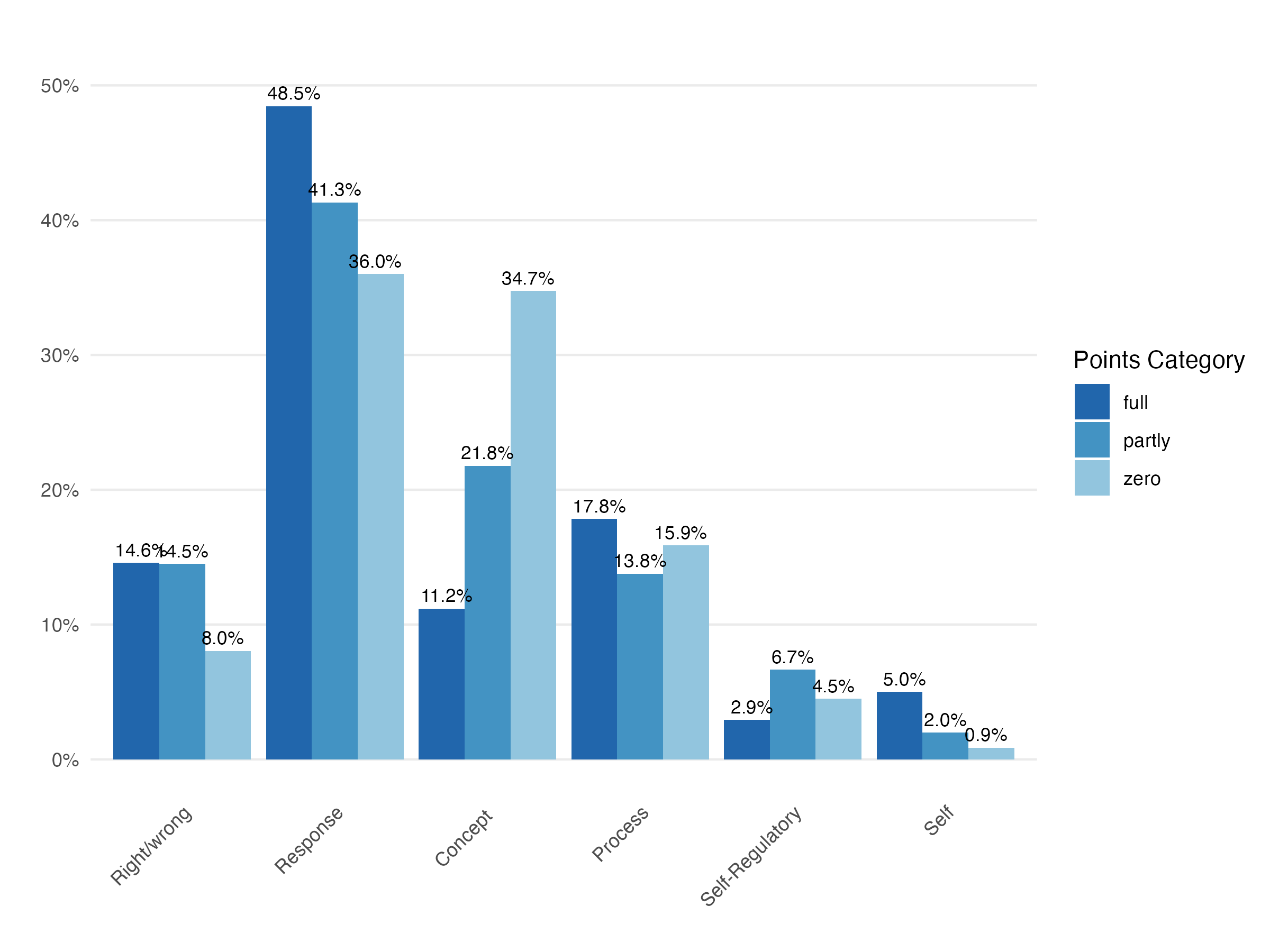}
    \caption{Coverage of feedback categories by points achieved.}
    \label{fig:coverage-points}
\end{figure}

The same pattern is visible for the coverage of feedback based on the points achieved (see fig. \ref{fig:coverage-points}:  Students that perform better receive less "conceptually-focused" feedback. In contrast to the quite stable occurrence rates described above, "response-oriented" feedback shows a different pattern  in coverage: Students achieving higher points receive more "response-oriented" feedback than students performing worse. That seems counter-intuitive since one would assume that wrong answers typically need more clarification on why they are wrong.

\subsubsection{Feedback structure by same points achieved}
Another important aspect of examining the structure of LLM-generated feedback is the consistency of the feedback across different students who received the same points. Such differences may lead to unequal learning opportunities if students receive less information despite comparable performance. To analyze this, we qualitatively examine question 1a. As this is a multiple-choice question with pre-defined answer options, it affords the highest degree of comparability. In this task, students were supposed to select all measures of central tendency (mean, median, mode), excluding the fourth answer option, standard deviation. We will begin by examining the feedback provided to students who answered the question correctly, followed by an analysis of the feedback for those who made errors.

We observed substantial differences in feedback even among students who all answered this question correctly. As highlighted by the exemplary feedback in Table \ref{tab:1a-full}, the first obvious difference is the length of the different feedback instances. Examples 1 and 2 both contain the same feedback categories: "right/wrong", "response-oriented" and "self" feedback. However, while Example 1 merely states that all selected answer options are correct because they are measures of central tendency, Example 2 adds another response-oriented piece of information: that the standard deviation, which was correctly omitted by the student, is a measure of dispersion. The student who received Example 1 did not get this information. 

Example 3 shows an even starker contrast. While it includes the additional information on the incorrect answer option, similar to Example 2, it also adds "conceptually-focused" feedback. This example shows how naturally feedback can progress: First stating that the task was solved correctly ("right/wrong"), then specifying why the answer was correct or incorrect ("response-oriented"), and adding a short explanation of the concepts involved ("conceptually-focused"). While we might assume based on their correct answers that these students \textit{know} which measures belong to which statistical concept, a brief reminder on their definition and what they are used for has the potential to further solidify their understanding.

\begin{table}[H]
    \centering
    \renewcommand{\arraystretch}{1.5}
    \begin{tabular}{c!{\vrule width 1pt}p{13cm}}
        \textbf{\#} & \textbf{Example feedback text} \\
        \Xhline{1pt}
        1 & {\sethlcolor{selfyellow}\hl{Very good}}, {\sethlcolor{responseblue}\hl{those are indeed all the measures of central tendency: the arithmetic mean (a), the median (b), and the mode (d).}} {\sethlcolor{rwblue}\hl{You solved this task correctly.}} {\sethlcolor{selfyellow}\hl{Keep it up!}} \\
        \hline
        2 & {\sethlcolor{rwblue}\hl{You solved the task correctly.}} {\sethlcolor{responseblue}\hl{The measures of central tendency in statistical analysis are indeed the arithmetic mean (a), the median (b), and the mode (d). The standard deviation (c), on the other hand, is a measure of dispersion.}} {\sethlcolor{selfyellow}\hl{Keep it up!}} \\
        \hline
        3 & {\sethlcolor{rwblue}\hl{You correctly identified the measures of central tendency.}} {\sethlcolor{responseblue}\hl{The arithmetic mean, the median, and the mode are indeed measures of central tendency.}} {\sethlcolor{conceptpurple}\hl{They give us an indication of where the "center" of a dataset is located.}} {\sethlcolor{responseblue}\hl{The standard deviation, which you correctly omitted, on the other hand, is a measure of dispersion,}} {\sethlcolor{conceptpurple}\hl{telling us how far the data are on average from the mean.}} {\sethlcolor{selfyellow}\hl{Keep it up!}} \\
    \end{tabular}
    \caption{English translations of selected feedback for task 1a (fully correct answers), highlighted by feedback category: "right/wrong" (\sethlcolor{rwblue}\hl{light blue}), "response-oriented" (\sethlcolor{responseblue}\hl{darker blue}), "conceptually-focused" (\sethlcolor{conceptpurple}\hl{purple}), "process" (\sethlcolor{processred}\hl{red}), "self-regulatory" (\sethlcolor{selfgreen}\hl{green}), "self" (\sethlcolor{selfyellow}\hl{yellow}).}
    \label{tab:1a-full}
\end{table}

Table \ref{tab:1a-partial} presents feedback responses for students who received partial points for task 1. In each case, students correctly identified “arithmetic mean” and “median” as measures of central tendency but omitted “mode.” This feedback is substantially longer than that for correct answers in Table \ref{tab:1a-full}, which is expected given the greater need for feedback when task-related knowledge gaps are determined. However, even among students with the same score, there are notable differences in the structure and content of feedback provided.

In Example 1, the student primarily receives “right/wrong” and “response-oriented” feedback, with the latter dominating. Two “self-regulatory” remarks are included, reminding the student to review all options carefully and to consider that multiple answers can be correct.
By contrast, Examples 2 and 3 instead include brief elements of “process” feedback. In Example 3, this feedback assesses the student's current learning status and encourages them to remember that “mode” is also a measure of central tendency. In Example 2, the feedback advises the student to revisit both concepts (measures of central tendency and dispersion) to deepen their understanding.

The biggest difference is the inclusion of “conceptually-focused” feedback,  absent in Example 1 but prominent in Examples 2 and 3, matching the share of “response-oriented” feedback in Example 1. This shift is meaningful: instead of simply giving feedback on the answer options, the feedback provides explanations of the underlying statistical concepts. 

Example 2 also identifies the incorrect answer (option 3) as a measure of dispersion and defines this term. In contrast, Example 3 omits this identification but offers definitions for all three measures of central tendency. This greater detail has the potential to highlight why these measures all belong to the same category, fostering a deeper understanding.

\begin{table}[H]
\centering
\renewcommand{\arraystretch}{1.5}
\begin{tabular}{c!{\vrule width 1pt}p{13cm}}
\textbf{\#} & \textbf{Example feedback text} \\
\Xhline{1pt}
1 & 
{\sethlcolor{rwblue}\hl{Your answer is almost correct, but not complete.}} 
{\sethlcolor{responseblue}\hl{You correctly identified that the arithmetic mean (a) and the median (b) are measures of central tendency. But you missed the mode (d), which is also a measure of central tendency. The standard deviation (c) is a measure of dispersion, not of central tendency, which you correctly excluded.}} 
{\sethlcolor{selfgreen}\hl{Always check all options carefully and remember that there can be multiple correct answers.}} 
{\sethlcolor{selfyellow}\hl{Good work}} 
{\sethlcolor{selfgreen}\hl{but don’t forget that your answer is only complete when it addresses all parts of the question.}} \\
\hline
2 & 
{\sethlcolor{rwblue}\hl{You already named two of the correct measures of central tendency}} 
{\sethlcolor{responseblue}\hl{– the arithmetic mean (a) and the median (b). However, you overlooked the mode (d), which is also a measure of central tendency.}} 
{\sethlcolor{conceptpurple}\hl{A small tip for the future: Measures of central tendency are statistical values that represent the central point of a distribution. These include not only the arithmetic mean and the median but also the mode. The standard deviation, on the other hand, is a measure of dispersion and indicates how much values deviate from the mean.}} 
{\sethlcolor{processred}\hl{Please review these concepts to develop a better understanding of the fundamental principles of statistics.}} \\
\hline
3 & 
{\sethlcolor{responseblue}\hl{You forgot the mode (d) in your answer, which is also a measure of central tendency.}} 
{\sethlcolor{processred}\hl{It appears you have a basic understanding of statistical measures of central tendency, since you correctly identified the arithmetic mean and the median.}} 
{\sethlcolor{conceptpurple}\hl{Measures of central tendency are statistical indicators that give us an idea of where the “center” of a distribution lies. They include the arithmetic mean, the median, and the mode. Each of these provides a different definition of “center.” • The arithmetic mean is the average of the values. • The median is the middle value when all values are arranged in ascending order. • The mode is the value that occurs most frequently.}} 
{\sethlcolor{processred}\hl{Please make sure to include the mode as a measure of central tendency in your future answers.}\tablefootnote{This "process" categorization is an example for the fringe cases that are expected with basing the categories on \cite{hattie_power_2007} where the levels are not completely disjunct but rather interrelated. In this case, the reminder to include it in the future answers can be interpreted as "self-regulatory", but as it rather hints towards reviewing the mode as a measure of central tendency and, hence, suggesting how to approach this statistical measure, it was labeled as "process" level. This can be compared with the rather different "self-regulatory" feedback in the first example of the table.}} \\
\end{tabular}
\caption{English translations of feedback for task 1a (partially correct answers), highlighted by feedback category: "right/wrong" (\sethlcolor{rwblue}\hl{light blue}), "response-oriented" (\sethlcolor{responseblue}\hl{darker blue}), "conceptually-focused" (\sethlcolor{conceptpurple}\hl{purple}), "process" (\sethlcolor{processred}\hl{red}), "self-regulatory" (\sethlcolor{selfgreen}\hl{green}), "self" (\sethlcolor{selfyellow}\hl{yellow}).}
\label{tab:1a-partial}
\end{table}

\section{Discussion and Conclusion}

Our results provide an in-depth analysis of LLM-generated feedback in a real (statistics) classroom setting with 70 students working on a mock exam. This specific course setting usually does not allow for individual feedback for every student. However, any errors or misleading responses could significantly hinder students’ learning progress. Subsequently, our approach was to use the text-generating capabilities of LLMs, while supplying the correct answers from the human instructors as contextual input for the LLMs to mitigate errors. To investigate if we can trust LLMs to deliver feedback of high quality in this context, we guided our study with two principal research questions:
\begin{enumerate}
    \item To what extent does the LLM produce errors while giving feedback when the instructors supply the correct answers as contextual input?
    \item What is the structure of the feedback generated, including its variations across tasks and students, and its alignment with established educational feedback theories?
\end{enumerate}

For our first research question, we found that 6.99\% (n=167) of the 2,389 feedback instances contained errors. Given the sensitivity of the setting, we applied a broad definition of errors, encompassing any occurrence potentially confusing, distracting, or misleading for students.  While a considerable number of these errors are simple one-word mistakes or erroneous translations, some include inaccurate or misleading explanations of statistical concepts. There were some tasks of the mock exam where errors occurred more frequently, but we found errors in the majority of the 38 task. Only 13 tasks contained no errors at all. We differentiated the most common errors into two categories based on their potential effect on students. Category 1 includes technical mistakes that are easily identifiable by students, such as a wrong handling of currencies (not accepting 0.30 Euro as a response for the correct answer 30 cents). While they still have the potential to be slightly disruptive and distracting, their negative impact is likely more limited as students can readily recognize them. On the other hand, Category 2 comprises more subtle issues, such as conceptually misleading explanations. These have a greater risk of introducing misconceptions without the student noticing, resulting in a ‘silent’ negative effect on learning. 

To address our second research question on feedback structure, we conducted a detailed content analysis of four exam tasks. Our analysis scheme is based on the established \cite{hattie_power_2007} framework, which we expanded by detailing the task-level feedback into three more subcategories drawn from \cite{ryan_beyond_2020}. We examined the structure of feedback both by frequency of occurrence and by text volume. Overall, we found expected outcomes such as "right/wrong" and "response-oriented" feedback being the most frequent across all students and tasks. Meaning, that in the vast majority of cases, the students got told if their answer is (in)correct ("right/wrong") and why ("response-oriented"). "Conceptually-focused" feedback, which explains the underlying concepts of a task, occurred less frequently but tended to be more elaborate when present. As these three categories are, in our approach, a dissection of the "task" level of \cite{hattie_power_2007}, it aligns strongly with previous empiric research that most feedback addresses the task level both for human and LLM contexts (\cite{dai_can_2023}). Our subdivision in three different types of the task level as the basis for a quantified analysis of qualitative data demonstrates that there can be more nuance to this feedback level. \citet{wisniewski_power_2020} showed that the impact of feedback is substantially influenced by the conveyed information content. This information content should rise strongly in our sub-categories from "right/wrong", to "response-oriented", and finally to "conceptually-focused".

“Process” feedback appeared in over half of the feedback instances, although often just as brief phrases reflecting the student’s current understanding and less often suggesting actual learning strategies. “Self-regulatory” feedback was by far the least frequent category, indicating students rarely received guidance on how to assess or monitor their own learning. This rare occurrence is consistent with previous quantifications of feedback levels \citep{dai_can_2023} and can mark a flaw in feedback design as this "self-regulatory" level can be considered as one of the most useful feedback levels (\cite{wisniewski_power_2020}. 

“Self” feedback, such as praise or personal comments (described as rarely effective by \cite{hattie_power_2007}) was the third most common category. However, it usually consisted of very brief remarks of praise (“Good job!”, "Great!"), resulting in the lowest overall text volume among all categories. The prevalence of such "self" feedback flattery aligns with evidence that LLMs can tend to over-optimize for human approval or validation at the expense of correctness or faithfulness, known under the term \textit{sycophancy} \citep{sharma_towards_2025}. This is amplified by cases in our feedback data where the LLM keeps flattering even when the student is not giving correct answers ("\textit{Good work}, but don’t forget that your answer is only complete when it addresses all parts of the question.") While this can be interpreted as encouragement, it could also be distracting from the actual suggestions that are given. \citet[p. 96]{hattie_power_2007} argued that praise "is unlikely to be effective, because it (...) too often deflects attention from the task." LLM studies indicate that \textit{sycophancy} is potentially a by-product of human-preference training, meaning that humans and, consequentially, preference models favor affirmative responses \citep{sharma_towards_2025}. This ambivalence is yet another connection to feedback research, where "self" feedback remains a complex and multi-layered topic. While in a recent revisit of the \citet{hattie_power_2007} model, "self" feedback was rated least useful by students \citep{mandouit_revisiting_2023}, other learner-centered research argues that "students still want to receive praise to facilitate their motivation moving forward" \citep[p. 3399] {van_boekel_feedback_2023}. 

We further analyzed feedback by the type of task (knowledge, interpretation, calculation), but found little interpretable patterns across these types. For example, one might expect interpretation tasks to elicit more “conceptually-focused” feedback, as they require deeper engagement with statistical concepts. Yet, one of the two interpretation tasks had the lowest frequency of this feedback type among the four analyzed.

Clearer patterns could be observed in the relation between feedback structure and the number of points achieved. Students receiving partial or no points received considerably more frequently “conceptually-focused” and “self-regulatory” feedback and less frequently “self” feedback. “Right/wrong” and “response-oriented” feedback remained relatively constant regardless of the performance. This is intuitive: students who answered correctly still received validation and praise, but less elaboration and guidance compared to those who struggled.

We also explored how consistent feedback was for students with similar performance. To do this, we qualitatively analyzed feedback for task 1a and found substantial variation, both among students who answered correctly and those who did not receive full points. In both groups, a key difference was whether or not students received “conceptually-focused” feedback. While this may be more negligible for students who answered correctly, it is potentially more problematic that students with identical incorrect answers sometimes received detailed explanations, while others did not.

Additionally, our qualitative analysis showed that even within the same category, feedback could differ strongly in content. For example, one feedback instance provided “conceptually-focused” explanations for both statistical concepts involved (central tendency and dispersion), while another addressed only central tendency, but elaborated in detail on the specific measures (mean, median, and mode). This indicates that even when the feedback structure is consistent, the actual content can vary considerably. 

Evaluating this feedback in depth presents a challenge, especially given its notable variance. The feedback levels of \cite{hattie_power_2007} are deliberately overlapping, as the most effective feedback should progress fluidly across levels. This requires judgment calls from coders, as some text segments may align with different levels depending on the perspective. Overall, our analysis shows that the levels, in combination with our extension of the task level based on the \cite{ryan_beyond_2020} categories, provide a suitable framework for accurately describing the structure and composition of feedback. 

Moreover, we believe this approach can serve as a foundation and guide for future feedback design. While many studies evaluate feedback (often drawing on \cite{hattie_power_2007}), they tend to focus on more elaborate tasks such as creative writing or project reports. These task types naturally provide more content and context for feedback than the compact, individual items used in our exam. Nonetheless, this particular context is critically important: in large-scale courses, students often receive no individual feedback at all. In this light, we must carefully weigh whether the presence of occasional errors or structural inconsistencies in LLM-generated feedback outweighs the alternative of providing no feedback whatsoever. One could also question how consistent instructor-generated feedback would actually be, and whether it would always avoid ambiguity or misinterpretation. Many instructors are not familiar with feedback theory and may not consistently provide the most effective and pedagogically sound comments. Still, it remains essential to hold automated systems to a higher standard, when the responsible human is no longer part of the feedback loop in order to intervene in case of problems.

Providing feedback via LLMs opens the door to offering individualized feedback where none would typically be available. In addition, it enables a shift from reactive, ad-hoc responses to a more deliberate, design-oriented approach to feedback. One can assume that instructors under common time constraints rarely design individualized feedback systematically and apply it consistently across all students. Instead, they might tend to comment on responses based on their teaching experience and personal didactic values. By leveraging LLMs, we gain the opportunity to invert this process: first defining what effective feedback should look like, and then deploying it consistently based on these predefined principles and frameworks. The framework presented in this paper can serve precisely as a foundation for such design. The key lies in making these feedback decisions consciously. For example, it may be appropriate to provide more detailed explanations to students who answered incorrectly than to those who answered correctly. But such decisions should be intentional, not arbitrary or inconsistent across similarly performing students. This engagement with deliberate feedback design also provides opportunities for instructors to sharpen exercise and exam design, ensuring that each task genuinely assesses the intended learning outcomes.

In this pilot study, we used the platform in an out-of-the-box configuration, with a generic prompt not specifically aligned with educational feedback theories. This setup reflects a realistic scenario in which instructors adopt such platforms with minimal customization. A promising avenue for future research is to evaluate how aligning prompts, incorporating retrieval-augmented generation (RAG), or applying fine-tuning with high-quality human feedback and our empirical findings affects the effectiveness of feedback. While RAG extends large language models by dynamically retrieving and integrating relevant external information into the generation process, fine‑tuning adapts the model’s parameters to domain‑specific data, thereby improving accuracy and contextual relevance. Exploring these approaches in combination with high‑quality human feedback could provide a more robust foundation for generating effective and pedagogically sound feedback.

Recent feedback research suggests to more strongly emphasize feedback literacy, meaning the learner's perception of feedback \citep{carless_development_2018, mandouit_revisiting_2023, weidlich_highly_2025}. We reflected this by conducting an evaluation survey: students responded in general very favorably to the implementation of the LLM-generated feedback. Particularly, 93 \% rated the feedback rather or very useful. Further, our survey data indicates that most students already use ChatGPT or other LLMs on their own for studying. Introducing it as part of an educational tool in the classroom with the opportunity to answer questions and address more common issues, provides an additional security net compared to students using those tools autonomously without supervision. In our error analysis, we mainly took the interpretation from the students' side, contemplating the weight of the errors by how easily students could detect them. Future work could deepen this literacy focus by embedding a feedback loop in which students review each feedback response, flagging potential errors and rating the helpfulness of the feedback. This would also align with the argument that feedback should be an ongoing dialogue \citep{mandouit_revisiting_2023}. The \textit{StudyLabs} platform offered students the opportunity to keep chatting with the feedback they received for every single task. This feature was very rarely used by the students, but could be a potentially valuable addition if used more consequentially and intentionally.

The present paper is not only a theory-informed technical and qualitative evaluation of AI-generated feedback in an authentic classroom setting. The feedback evaluation framework that we extended from \cite{hattie_power_2007} in combination with \cite{ryan_beyond_2020} is not limited to the assessment of LLM feedback, it is equally applicable to human-generated feedback. Through this combined framework, we conducted a fine-grained analysis that dissects common task-level feedback in greater detail and offers new insight into feedback composition. This demonstrates how engaging with LLM-generated feedback goes beyond merely evaluating the implementation of new AI systems, it opens up broader discussions about the nature, quality, and design of educational feedback at its core.

\section*{Declarations}
\textbf{Availability of data and materials} \\
We published the anonymized pre- and post-survey data, analysis code and (translated) codebooks: \href{https://osf.io/z9wum/?view_only=4e9ae3df5fc94edca27876fe3a44f712}{Link to OSF repository}.
The repository is currently published anonymously for blinded review. Please note that we published the GPT-usage and StudyLabs-experience data from the survey (as reported in Results Section 4.1), while not publishing sociodemographic information, study program data, and open-text responses to prevent potential re-identification through variable combinations.

We attached a translation of the mock exam including task type categorizations as part of this submission as supplementary material.

Regarding the potential publication of the feedback data from the StudyLabs platform (including scoring and feedback) and our subsequent categorization data, we are still in contact with ZAVI as the data owners. We will update this statement when a decision is reached.

\textbf{Funding} \\
This research did not receive any specific grant from funding agencies in the public, commercial, or not-for-profit sectors.

\textbf{Generative AI and AI-assisted technologies in the writing process}\\
Generative AI (genAI) in the form of Grammarly and ChatGPT was used for language improvements. Further, ChatGPT was used for assistance in translation, literature discovery, R programming, and LaTeX coding. Generative AI tools were \textbf{not} used at any stage to process raw data or draft the manuscript.

\textbf{Acknowledgments} \\
We thank \textit{ZAVI} for providing the software as well as the data.

\bibliography{refs-studylabs}

\appendix
\pagestyle{plain}
\section{Appendix} 
\subsection{Most common feedback errors}
\begin{landscape}
\thispagestyle{plain}
{\setlength{\tabcolsep}{3pt} 
\footnotesize
\begin{longtable}{@{}P{.07\linewidth}P{.04\linewidth}P{.24\linewidth}P{.65\linewidth}@{}}
\caption{Most common feedback errors with example text (translated)}
\label{tab:errors-most-common}\\
\toprule
\textbf{Task} & \textbf{\#} & \textbf{Error} & \textbf{Example (translated from German)}\\
\midrule
\endfirsthead
\toprule
\textbf{Task} & \textbf{\#} & \textbf{Error} & \textbf{Example (translated from German)}\\
\midrule
\endhead
\bottomrule
\endfoot

7a & 10 & Does not accept ``0.3021 Euro'' for ``30 cent'' & 
``You correctly recognized the concept of the estimated coefficient in this context, but in this case there seems to have been a small mix-up. The interpretation of the coefficient 0.3021 actually means an increase of 0.3021 Euro (about 30 cents) per one-unit increase in the predictor, not ``30 cent'' as a separate unit. Clarifying this helps ensure a correct understanding of the matter. (…)"\\[0.6em]

3f & 34 & Ignores new table for the exercise, refers to the previous exercise description instead & 
``Although your calculation is basically correct, you apparently used the wrong table. In the exercise prompt, the values 23, 41, 20, and 13 are given. (…)"\\[0.6em]

6b & 14 & Shows correct formula for calculating $\chi^2$ from the provided data but explains a different formula (first computing expected values) & 
``(…) The $\chi^2$ calculation consists of several steps that sum $(\text{observed}-\text{expected})^2$ divided by the expected frequency across all cells. In this case, with the given data, the calculation would be: 
$$
\chi^2 = 3467 \cdot \Big( \frac{191^2}{397\cdot 1760} + \cdots + \frac{1120^2}{2374\cdot 1707} - 1 \Big)
$$
This yields a value of $13.59$ for the $\chi^2$ statistic. (…)"\\[0.6em]

1e & 6 & Missing distinction between Spearman and Pearson correlations (creates the impression that Spearman can only be used for metric variables) & 
``(…) The Pearson correlation (answer b) and the Spearman correlation (answer a), on the other hand, measure the relationship between two continuous variables and are therefore not suitable in this case."\\[0.6em]

3c, 3d & 8 & Misleading confusion between ordinal and nominal variables & 
``(…) Regardless of whether the data were collected on a nominal or an ordinal scale, the median can be estimated as long as the data can be ordered. (…)"\\[0.6em]

7h & 30 & Assumes the wrong scale of the independent variable (variable describes the dichotomous existence of children, not their number) & 
``(…)
1) Identify the independent variables. In this case they are income and the number of children in the household. 2) Identify the dependent variable. This is the amount a household spends on a summer vacation. 3) Use the given model for prediction. If you plug in the values, you obtain the predicted expenses for the vacation. 
Now take the given values and plug them into the formula: income $= 2700$ Euro and children $= 2$. This yields the following calculation:
$$259.02195 + 0.27891 \cdot 2700 - 66.71902 \cdot 2 = 945.3599.$$ (…)"\\

\end{longtable}
}
\label{tab:errors-common}
\end{landscape}

\subsection{Abbreviations}

\begin{tabular}{@{}l@{\hspace{2cm}}l@{}}
AI   & Artificial Intelligence \\
e.g. & exempli gratia \\
i.e. & id est \\
GPT  & Generative Pre-trained Transformer \\
LLM  & Large Language Model
\end{tabular}

\end{document}